\UseRawInputEncoding 

\documentclass[twocolumn,showpacs,floats,floatfix,superscriptaddress,aps,pra]{revtex4-1}

\usepackage{amsfonts}
\usepackage{amssymb}
\usepackage{amsmath}
\usepackage{calc}
\usepackage{graphicx}
\usepackage{bm}

\usepackage[normalem]{ulem}
\usepackage{amsmath,amssymb}
\usepackage{multirow}
\usepackage{xcolor,soul}
\usepackage{srcltx}
\usepackage{hyperref,graphicx}

\def\be{ \begin{equation}}
\def\ee{ \end{equation}}
\def\bea{ \begin{eqnarray}}
\def\eea{ \end{eqnarray}}
\def\bse{ \begin{subequations}}
\def\ese{ \end{subequations}}
\def\bc{ \begin{center}}
\def\ec{ \end{center}}

\begin{document}

\author{Stefano Longhi$^{*}$} 
\affiliation{Dipartimento di Fisica, Politecnico di Milano, Piazza L. da Vinci 32, I-20133 Milano, Italy}
\affiliation{IFISC (UIB-CSIC), Instituto de Fisica Interdisciplinar y Sistemas Complejos, E-07122 Palma de Mallorca, Spain}
\email{stefano.longhi@polimi.it}

\title{Non-Hermitian skin effect beyond the tight-binding models}
  \normalsize


%
\bigskip
\begin{abstract}
\noindent  

The energy bands of non-Hermitian systems
exhibit nontrivial topological features that arise from the complex nature of the energy spectrum.  Under 
periodic boundary conditions (PBC), the energy spectrum describes rather generally  closed loops in complex plane, characterized by integer nonzero winding numbers.
Such nontrivial winding provides the topological signature of the non-Hermitian skin effect (NHSE), i.e. the macroscopic condensation of bulk states at the lattice edges under open boundary conditions (OBC). 
In spite of the great relevance of band winding in the non-Hermitian topological band theory and the related NHSE, most of current results rely on tight-binding models of non-Hermitian systems, while exact Bloch wave function analysis of the NHSE and related topological band theory is still lacking. While tight-binding models can correctly describe narrow-band electronic states with a relatively weak degree non-Hermiticity, they are not suited to describe high-energy wide-band electronic states and/or regimes corresponding to strong non-Hermiticity. Here we consider the single-particle continuous Schr\"odinger equation in a periodic potential, in which non-Hermiticity is introduced by an imaginary vector potential in the equation, and show that the NHSE is ubiquitous under OBC and characterized by a non-vanishing integer winding number, even thought the energy spectrum under PBC always comprises an open curve, corresponding to high-energy electronic states. We also show that the interior of the PBC energy spectrum corresponds to the complex eigenenergies sustaining localized (edge) states under semi-infinite boundary conditions. 
 \end{abstract}

\maketitle

\section{Introduction}
A common belief in solid state physics is that bulk physics 
is insensitive to boundary conditions, so that bulk properties such as 
thermodynamic and transport
quantities \cite{Ziman,Atland} as well as band topological invariants \cite{Kane} can be 
computed assuming the Born-von Karman (periodic) boundary conditions. 
This common wisdom has been challenged in recent years, 
where effective non-Hermitian Hamiltonians describing open lattice systems 
display strong sensitivity to boundary conditions and richer non-Hermitian topology arising from the complex nature of the energy spectrum \cite{r1,r2,r3,r4,r5,r6,r7,r8,r9,r10,r11,r12,r13,r14,r15,r16,r17,r18,r19,r20,r21,r22,r23,r24,r25,r26,r27,r28,r29,r30,r31,r32,r33,r34,r35,r36,r37,r38,r39,r40,r41,r42,r43,r44,r45,r46,r47,r48,r49,r50,r51,r52,r52b,r53,r54,r56,r57,r58,r59,r60,r61,r62,r63,r64,r65,r66,r67} (for recent reviews see \cite{r11,r45,r47}). Unlike Hermitian lattices, non-Hermitian ones can be topologically nontrivial even in one dimension and without any symmetry 
because the energy spectrum can form closed loops in the complex
plane, characterized by a non-vanishing winding number \cite{r4,r35,r47,r49}. Such a nontrivial winding provides the topological
grounds of exotic phenomena observed in non-Hermitian lattices, such
as the breakdown of the bulk-boundary correspondence based on Bloch band topological invariants and the non-Hermitian skin effect (NHSE) \cite{r47}.
Generalized Brillouin zone and non-Bloch band theory have been formulated to correctly predict the topological edge modes from the topological bulk invariants, which are defined in the generalized Brillouin zone rather than in the standard Brillouin zone \cite{r7,r17,r34}. The point-gap topology of the energy spectrum under periodic boundary conditions (PBC), corresponding to a non-vanishing winding number, is at the heart of the NHSE, i.e. to the macroscopic condensation of bulk modes at the edges in a lattice under open boundary conditions (OBC) \cite{r31,r35,r49}. A central result relating non-Hermitian point-gap topology and NHSE is that, whenever the energy spectrum under PBC describes a closed loop in complex plane with a non-vanishing winding number, under the OBC the energy spectrum of the same system collapses to an open curve with trivial topology in the interior of the  PBC energy spectrum, and correspondingly the NHSE is observed, i.e. the wave functions are squeezed toward the edges of the lattice \cite{r35,r49}.  A paradigmatic model exhibiting the NHSE was introduced by Hatano and Nelson more than two decades ago \cite{Nelson1,Nelson2,Nelson3} as a non-Hermitian extension of the famous Anderson model of localization. In their model, Hatano and Nelson introduced an imaginary vector potential in the non-relativistic single-particle Schr\"odinger equation \cite{S1,S2}. When the  Schr\"odinger equation describes a quantum particle on a one-dimensional periodic potential, within a tight-binding model the Peierls phase in the hopping amplitudes introduced by the imaginary vector potential takes the form of a phase factor but with a real exponent \cite{S1}, i.e. it induces asymmetric left/right hopping along the lattice responsible for the NHSE. The implications of the imaginary gauge field in the Anderson-Hatano-Nelson model have been discussed in several subsequent works (see e.g. \cite{sub1,sub2,sub3,sub4,sub5,sub6,sub7,sub8,sub9,sub10,sub11,sub12,sub13}) and found interesting applications in robust excitation transport \cite{sub10,sub11} and laser array stabilization \cite{sub12,sub13,sub14}. The point-gap topology of the Hatano-Nelson model was disclosed in Ref.\cite{r4}, which inspired much of the current studies on non-Hermitian topological models displaying the NHSE. \par All such previous studies on the NHSE and related point-gap topology have been concerned with the tight-binding approximation of lattice bands, which is suitable to describe low-energy narrow bands separated by wide gaps. However, it is well known that wide  energy bands in a crystal separated by narrow gaps, corresponding to the nearly-free electron limit, cannot be described by the tight-binding model, and that both the nearly-free electron model and the tight-binding model are approximate methods to describe the band structure of a crystal (Fig.1).
\begin{figure}[t]
   \centering
    \includegraphics[width=0.45\textwidth]{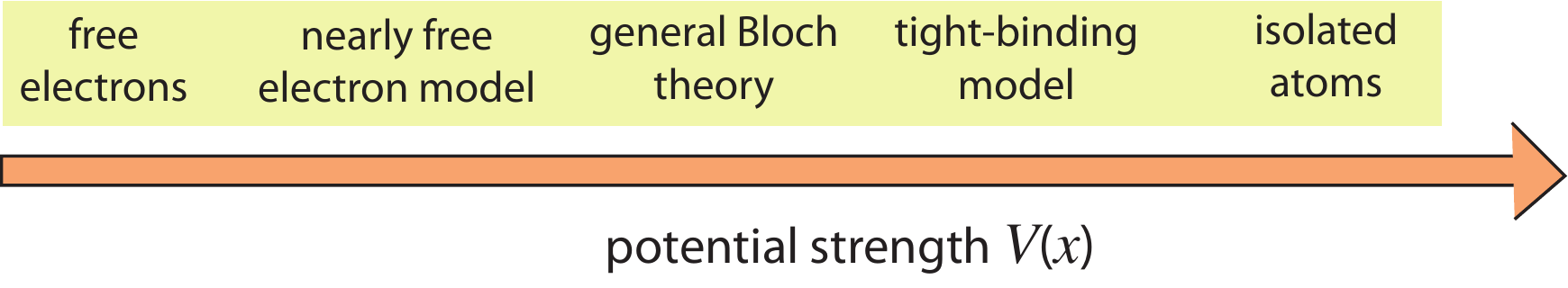}
    \caption{Schematic of the models used to describe the band structure of a crystal as a function of the potential strength.}
     \label{fig1}
\end{figure}
 An open question is whether the NSHE and related point-gap topology of energy bands can be extended beyond the tight-binding models, i.e. within the general Bloch-Floquet theory of periodic potentials. \par
In this work we consider the single-particle Schr\"odinger equation in a periodic potential, in which non-Hermiticity is introduced by an imaginary vector potential in the equation \cite{S1,S2}. We show that the energy spectrum is strongly sensitive to the boundary conditions, and that under OBC one observes the NHSE, with bulk modes squeezed toward the edge of the crystal. Correspondingly, the band structure displays a point-gap topology with non-vanishing winding of all bands. However, contrary to the tight-binding models, the energy spectrum of the highest energy band of the crystal describes an open (rather than a closed) curve in complex energy plane, which is nevertheless still characterized by a non-vanishing integer winding number. As the imaginary gauge field is increased, band merging is observed, until a single energy band emerges, described by an open curve in complex energy plane approaching  the energy dispersion curve of the free-particle limit (a parabola). The results are illustrated considering three significant examples of potentials, namely the Lam\'e potential, the binary (double-well) potential, and the Mathieu (sinusoidal) potential.

\section{Energy spectrum, non-Hermitian skin effect and winding number in a crystal with an imaginary gauge field}
The starting point of our analysis is provided by the Schr\"odinger equation for a quantum particle in a periodic one-dimensional potential $V(x+a)=V(x)$ with lattice period $a$ and with an imaginary gauge field $\beta$ \cite{S1,S2}. In scaled form, the  Schr\"odinger equation reads
\begin{equation}
E \psi(x)= -(\partial_x+\beta)^2 \psi (x)+V(x) \psi (x) \equiv \hat{H}_{\beta} \psi (x). \label{schroed}
\end{equation}
For the sake of definiteness, in the following we will assume $\beta \geq 0$, however the results are valid {\it mutatis mutandis} also for the $\beta \leq 0$ case.
Let us consider a crystal made of $M$ unit cells with either OBC (infinite edge barriers)
\begin{equation}
\psi(0)=\psi(Ma)=0 \label{OBC}
\end{equation}
or PBC
\begin{equation}
\psi(Ma)=\psi(0) \label{PBC}
\end{equation}
in the larger $M$ limit. A semi-infinite crystal on the line $x \geq 0$ can be also considered, corresponding to the semi-infinite boundary conditions (SIBC)
\begin{equation}
\psi(0)=0 \; , \; \; {\rm{max}} \lim_{x \rightarrow + \infty} |\psi(x)|< \infty. \label{semiinf}
\end{equation}
While in the Hermitian limit $\beta=0$ the energy spectrum of $\hat{H}_{\beta}$ does not substantially depend on the boundary conditions (either OBC, PBC or SIBC), a strong dependence of energy spectrum on boundary conditions is found in the non-Hermitian regime $\beta \neq 0$.
\subsection{Energy spectrum under OBC and the skin effect}
For OBC, after the imaginary gauge transformation
\begin{equation}
\psi(x)=\phi(x) \exp(- \beta x) \label{gauge}
\end{equation}
Eq.(\ref{schroed}) reads
\begin{equation}
E \phi(x)= -\partial_x^2 \phi (x)+V(x) \phi (x)= \hat{H}_{\beta=0} \phi(x) \label{schrodgauge}
\end{equation}
with
\begin{equation}
\phi(Na)=\phi(0)=0 \label{OBCgauge}.
\end{equation}
This means that $\phi(x)$ are the usual extended Bloch wave functions in a Hermitian crystal of finite length with energy spectrum defined by the Bloch bands of the infinite crystal, with possible additional surface Tamm states localized at the edges of the lattice \cite{Tamm1,Tamm2,Tamm3}.
Such spectrum is entirely real and is described by a finite set (or an infinite numerable set) of intervals $I_1$, $I_2$, ..., $I_n$, ... on the real energy axis of permitted energies (bands) separated by intervals of forbidden energies (gaps). The energy spectrum is bounded below but not above. This means that, for a crystal with a finite number $(N-1)$ of gaps and supporting $N$ bands, the interval $I_N$ of the highest energy band is a semi-infinite line on the real energy axis, extending to $E=+\infty$ [see Fig.2(a)].
Hence under OBC the imaginary gauge field $\beta$ does not change the energy spectrum of $H_{\beta}$ as compared to the Hermitian limit $\beta=0$. {{However, the imaginary gauge field changes the localization properties of the wave functions and is responsible for the appearance of the NHSE under OBC. In fact, according to 
Eq.(\ref{gauge}) all extended Bloch wave functions $\phi(x)$ in the Hermitian limit become exponentially localized toward the left or right edge of the lattice (depending on the sign of $\beta$), with a localization length $\sim 1 / | \beta|$.} Also, since the energy spectrum remains entirely real and formed by a set of straight open curves, it is topological trivial.}
Such results clearly indicate that the appearance of the NHSE by an imaginary gauge field in a finite crystal with OBC is a very general feature, that holds beyond the usual tight-binding models of the crystal.
\subsection{Energy spectrum under PBC}
Under PBC, the imaginary gauge transformation (\ref{gauge}) cannot be used to eliminate the imaginary field $\beta$, and the energy spectrum of $\hat{H}_{\beta}$ should be computed rather generally as follows. 
Indicating by $k$ the Bloch wave number in the first Brillouin zone ($- \pi/a \leq k < \pi/a$), the PBC  Eq.(\ref{PBC}) is satisfied by letting
\begin{equation}
\psi(x)= \exp(i kx) \sum_n \psi_n \exp( i2 \pi i n x/a ) \label{Bloch}
\end{equation}
\begin{figure}[t]
   \centering
    \includegraphics[width=0.45\textwidth]{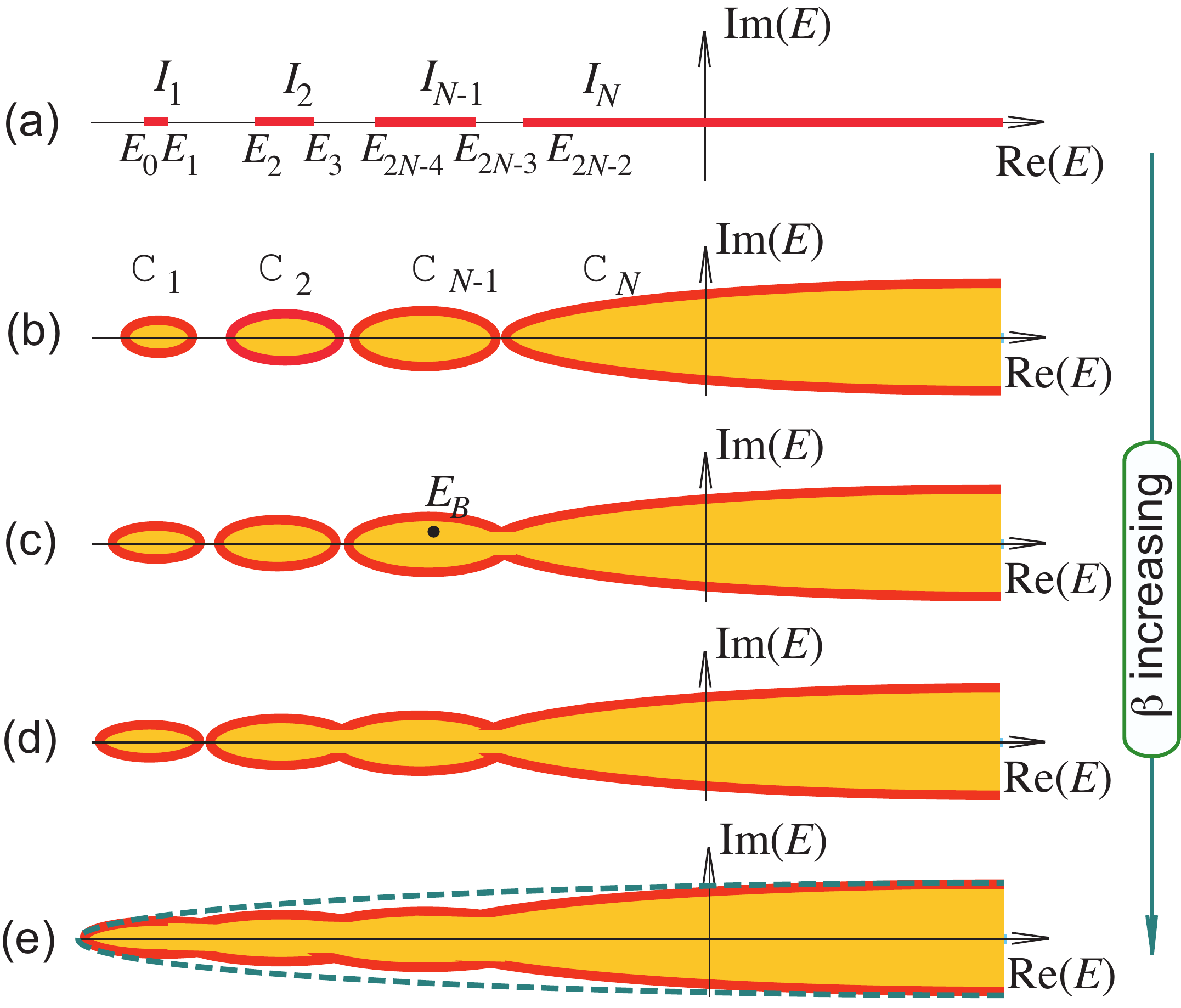}
    \caption{Schematic of the energy spectrum in complex plane $E$ (solid red curves) of the non-Hermitian Hamiltonian $\hat{H}_{\beta}$ under PBC for increasing values of the imaginary gauge field $\beta$. (a) $\beta=0$ (Hermitian limit). For a crystal with a finite number $(N-1)$ of gaps, the energy spectrum comprises $N$ intervals (energy bands) $I_1$, $I_2$,..., $I_N$ on the real energy axis, separated by $(N-1)$ gaps, with $I_{N}$ unbounded from above. The $(2N-1)$ band edges are denoted by $E_0$, $E_1$, ..., $E_{2N-2}$. (b) As the gauge field $\beta$ is slightly increased above zero, the energy spectrum is described by $(N-1)$ non-intersecting closed loops $\mathcal{C}_1$, $\mathcal{C}_2$, ..., $\mathcal{C}_{N-1}$, emanating from the straight segments $I_1$, $I_2$,..., $I_{N-1}$, and an additional open curve $\mathcal{C}_N$ emanating from the semi-infinite line $I_N$. (c-e) As the imaginary gauge field $\beta$ is further increased, successive merging of the curves $\mathcal{C}_n$ is observed, thus reducing the number of closed curves in complex energy plane. Above a critical value $\beta=\beta_c$, all bands merge and the energy spectrum is described by an open curve [panel (e)], which is approximated by the parabola (dashed curve) of the free-particle problem as $\beta \rightarrow \infty$. The shaded areas internal to the various bands correspond to the domain of base energies $E_B$ where the winding $W(E_B)$ is non vanishing and edge states do exist in the semi-infinite lattice.}
     \label{fig1}
\end{figure}
with quantized $k$ ($kMa$ should be an integer multiple than $ 2 \pi$; in the large $M$ limit $k$ can be taken as a continuous variable varying in the first Brillouin zone). {{Note that under PBC the wave functions $\psi(x)$, as given by Eq.(8), are extended states over the entire lattice, i.e. they are of Bloch type, rather than being squeezed toward the edges as in a system with OBC.}}
The crystal energy bands $E_{\beta}(k)$ for $\beta \neq 0$ are obtained from the eigenvalue equation
\begin{eqnarray}
E_{\beta}(k) \psi_n & = & - (ik+ 2 \pi i n/a + \beta)^2 \psi_n + \sum_m V_{n-m} \psi_m \nonumber \\
& \equiv & \sum_m \mathcal{H}_{n-m}(k) \psi_l \label{eigenvalue}
\end{eqnarray}
where $V_n=(1/a) \int_0^a dx V(x) \exp(-2 \pi i n x /a)$ are the Fourier coefficients of the periodic potential $V(x)$. 
Equation (\ref{eigenvalue}) clearly indicates that the energy spectrum in the non-Hermitian regime $\beta \neq 0$ is obtained from the spectrum in the Hermtian limit $\beta=0$ by the simple relation
\begin{equation}
E_{\beta}(k)=E_{\beta=0}(k-i \beta) \label{complexification}
\end{equation}
i.e. after complexification of the Bloch wave number $k$ ($k \rightarrow k-i \beta$). { This result is somehow analogous to the one found in the tight-binding limit, where complexification of $k$ corresponds to the introduction of a generalized Brillouin zone  to describe the energy spectrum of the non-Bloch Hamiltonian under OBC \cite{r7,r10,r17}. However, note that in our case Eq.(10) relates the energy spectra of two distinct Bloch Hamiltonians, one with $\beta=0$ and the other one with $\beta \neq 0$, and PBC are assumed in both cases. Interestingly, in the Bloch (momentum) space the matrix Hamiltonian $\mathcal{H}(k)$ defined by Eq.(9) satisfies the symmetry $\mathcal{P} \mathcal{T} \mathcal{H}(k)= \mathcal{H}(-k) \mathcal{P} \mathcal{T}$, with parity $\mathcal{P}$ and time-reversal $\mathcal{T}$ operators defined by $\mathcal{P} \psi_n=\psi_{-n}$ and $\mathcal{T}=\mathcal{K}$ ($\mathcal{K}$ is the element wise complex conjugation operation). Therefore, as $k $ spans the Brillouin zone, the eigenenergies of the $\hat{H}$ appear is complex-conjugate pairs.}
In the Hermitian limit $\beta=0$, the energy spectrum under PBC coincides with the one under OBC, with the exception of possible isolated energies corresponding to edge (surface) Tamm states. 
As the imaginary gauge $\beta$ is increased from zero, i.e. after complexification of the Bloch wave number $k$, under PBC each band undergoes a continuous reshaping and describes rather generally a closed curve $\mathcal{C}_n$ in complex energy plane, emanating from the corresponding straight segment $I_n$ on the real energy axis at $\beta=0$ [Fig.2(b)]. Such a result simply follows from the Fourier form of the dispersion curve of each band in the Hermitian limit and from the complexification of $k$ in the non-Hermitian regime [Eq.(\ref{complexification})]. An exception is provided by the curve $\mathcal{C}_N$ emanating from the highest-energy (continuum) band $I_N$, which describes an open curve $\mathcal{C}_N$. As shown in Sec.III and schematically illustrated in Figs.2(c), (d) and (e), successive band merging arises as $\beta$ is increased, until at large imaginary gauge fields only a single band, described by an open curve in complex energy plane, is observed [Fig.1(e)], which approaches the free-particle dispersion curve [$V=0$ in Eq.(\ref{schroed})]
\begin{equation}
E_{\beta}(\kappa) \simeq V_0 - (i \kappa+\beta)^2. \label{parabola}
\end{equation}
($-\infty < \kappa < \infty$) in the large $\beta$ limit. Note that  in such a limit the open curve is described by a parabola, depicted by a dashed curve in Fig.2(e). 
{{
At each band merging point, the underlying Hamiltonian becomes defective, corresponding to the appearance of an exceptional point or spectral singularity \cite{referee}. This result will be illustrated in Sec.III.C within a nearly-free electron model.\\
 Since under PBC the energy spectrum is described by a set of curves in complex energy planes, it can show a non-trivial topology characterized by a non-vanishing spectral winding number.
According to previous works  \cite{r4,r35,r49,r65}, for any point-gap (basis) energy $E_B$ one can define a winding number  of the PBC energy spectrum by the relation
\begin{equation}
W(E_B)= \sum \frac{1}{2 \pi i } \int_{-\pi}^{\pi} dk \frac{d}{dk} \log \left\{ E_{\beta}(k)-E_B \right\}
\end{equation}
where the sum is extended over the various bands of the crystal.} }Clearly, $W(E_B) \neq 0$ whenever the point-gap energy $E_B$ is in the interior of one of the closed curves $\mathcal{C}_n$ ($ n<N$), or on the right side of the open curve $\mathcal{C}_N$, as shown by the shaded areas in Fig.2(b). { We note that, even though the curve $\mathcal{C}_N$ is open, the winding number $W$ is still quantized. Intuitively, this follows from the fact that the curve closes sufficiently fast at infinity, i.e. $|Im(E) / Re(E)| \rightarrow 0$ as we move at infinity along the  curve $\mathcal{C}_N$.} 
As $\beta$ is increased and band merging occurs, a similar scenario is found, with $W(E_B) \neq 0$ when $E_B$ is chosen in the shaded areas internal to the distinct bands [Figs.2(c) and (d)].
In particular, in the large $\beta$ limit $|W(E_B)|=1$ for any base energy $E_B$ in the interior  of the parabolic curve, defined by Eq.(\ref{parabola}).\\

\subsection{Winding number, energy spectrum and edge states in the semi-infinite lattice}
A main result in non-Hermitian tight-binding models with a nontrivial point-gap topology is provided by Theorem I of Ref.\cite{r35}, that relates the interior of the closed loops describing the PBC energy spectrum with the energy spectrum of the system under SIBC. In Appendix A we show that such a main result is valid also in the continuous model. This means that any base energy $E_B$, such that $W(E_B )\neq 0$, does belong to the energy spectrum of $\hat{H}_{\beta}$ under SIBC, i.e. there exists a wave function (edge state) $\psi(x)$ to Eq.(\ref{schroed}) with eigenenergy $E=E_B$ and with
\begin{equation}
\psi(0)=0 \; ,  \; \; \lim_{x \rightarrow + \infty} \psi(x)=0. 
\end{equation}
 Moreover, as $E_B$ approaches the domain boundaries, i.e. $E_B$ lies on the lines $\mathcal{C}_1$, $\mathcal{C}_2$,..., $\mathcal{C}_N$, the wave function $\psi(x)$ becomes an extended state, thus still belonging to the spectrum of $\hat{H}_{\beta}$ under the SIBC (\ref{semiinf}). {It should be noted that the localized edge states satisfying Eq.(13) under SIBC should not be confused with the skin modes, which are observed under OBC in a finite lattice.}

\section{Illustrative examples}
In this section, we illustrate the general results presented in the previous section by considering a few examples of one-dimensional crystals with specific shape of potential $V(x)$ known in the literature, which cannot be fully described within a tight-binding model. We recall that in the tight-binding model the potential $V(x)$ is written as a periodic sequence of quantum wells $V_a(x)$ (atomic potentials)
\begin{equation}
V(x)=\sum_n V_a(x-na) \label{TBP}
\end{equation}
with negligible overlapping between adjacent potential wells in the lattice.
In the Hemitian limit $\beta=0$, a given bound state (atomic orbital) $u_a(x)$ of the potential well $V_a(x)$ gives rise to a tight-binding  band with Bloch wave functions written as a linear combination of atomic orbitals (LCAO)
\begin{equation}
\psi(x)=\sum_n \psi_n u_a(x-na) 
\end{equation}
with amplitudes $\psi_n$ satisfying the eigenvalue equation \cite{Atland}
\begin{equation}
E \psi_n= \sum_l \mathcal{H}_l  \psi_{n-l}.
\end{equation}
In the above equation, the hopping amplitudes $\mathcal{H}_l$ are given in terms of overlapping integrals
\begin{equation}
\mathcal{H}_l=\int dx\; u_a^*(x-la) \hat{H}_{\beta=0} \; u_a(x).
\end{equation}
The corresponding dispersion curve $E_{\beta=0}(k)$ of the tight-binding band, originating from the atomic orbital $u_a(x)$, is obtained from the Ansatz $\psi_n=\psi_0 \exp(i ka n)$ ($-\pi/a \leq k < \pi/a$) and reads
\begin{equation}
E_{\beta=0}(k)=\sum_l \mathcal{H}_l \exp(-ikla). \label{dispTB}
\end{equation}
In the non-Hermitian case $\beta \neq 0$, according to Eq.(\ref{complexification}) the dispersion curve $E_{\beta}(k)$ under PBC is obtained from Eq.(\ref{dispTB}) after the replacement $k \rightarrow k-i \beta$. Correspondingly, the PBC energy spectrum describes a closed loop in complex energy plane with a point-gap topology and associated non-vanishing winding number $W(E_B)$ for any base energy $E_B$ internal to the loop \cite{r35,r65}. In particular, the tight-binding Hatano-Nelson model \cite{S1} is obtained in the nearest-neighbor tight binding limit $\mathcal{H}_l \simeq 0$ for $l \neq 0, \pm 1$, corresponding to a sinusoidal dispersion curve $E_{\beta=0}(k)$ in the Hermitian limit and a curve $E_{\beta}(k)$ describing an ellipse in complex energy plane for $\beta \neq 0$ \cite{r4}.\\
While the tight-binding model is appropriate to describe narrow bands separated by wide gaps, or couple of mini bands separated by a small gap, it clearly fails to provide a correct analysis of wide bands separated by small gaps, and to describe the continuum band of high-energy electron states. Even for narrow bands separated by large gaps the tight-binding description becomes inadequate at large imaginary gauge fields $\beta$ owing to the phenomena of band merging and non-Hermiitan delocalization of the atomic (orbital) wave functions. Therefore, to correctly describe the energy spectrum under PBC one should resort to the general Bloch formulation of wave functions in the framework of the continuous Sch\"odinger equation.

\subsection{The Lam\'e potential}
The first illustrative example is provided by the Lam\'e potential, which is an exactly solvable model of periodic potential exhibiting a finite number of energy gaps. The Lam\'e potential reads \cite{Lame1,Lame2,Lame3,Lame4}
\begin{equation}
V(x)=N(N-1) \left\{ m \; {\rm sn}^2(x;m) -1 \right\} \label{Lame}
\end{equation}
where $N$ is a positive integer number ($N \geq 2$) and ${\rm sn}(x;m)$ is a Jacobi elliptic function of real elliptic modulus parameter $m$ ($0<m<1$), i.e. ${\rm sn}(x;m)= \sin \varphi$ with
\begin{equation}
x=\int_0^{\varphi} \frac{d \theta}{\sqrt{1-m \sin^2 \theta}}.
\end{equation}
The potential (\ref{Lame}) is periodic with period $a=2 K(m)$, where 
\begin{equation}
K(m)=\int_0^{\pi /2} \frac{d \theta}{\sqrt{1-m \sin^2 \theta}}.
\end{equation}
is the first complete elliptic integral. Interestingly, the Lam\'e potential has exactly $(N-1)$ gaps and  $N$ bands, as schematically shown in Fig.2(a). The $(2N-1)$ band edges
$E_0$, $E_1$, ..., $E_{2N-2}$ are algebraic functions of the parameter $m$, i.e.  they are the roots of a certain polynomial, the coefficients of which are polynomial in $m$ \cite{Lame4}. In particular, for $N=1$, i.e. for a crystal with a single gap, the band edges read
\begin{equation}
E_0=m-2 \; , \; \; E_1=-1 \; , \; \; E_2=-1+m. \label{Eedges}
\end{equation} 
In the limit $m \rightarrow 1ì$, the period $a$ diverges and, within each period, the Lam\'e potential $V(x)$ is very well approximated by the reflectionless P\"oschl-Teller potential well \cite{Lame1,PT1,PT2,PT3}, i.e. $V(x)$ can be written as in Eq.(\ref{TBP}) with
\begin{equation}
V_a(x)=-\frac{(N-1)N}{\cosh^2 x} \label{PTwell}
\end{equation}
\begin{figure}[t]
   \centering
    \includegraphics[width=0.47 \textwidth]{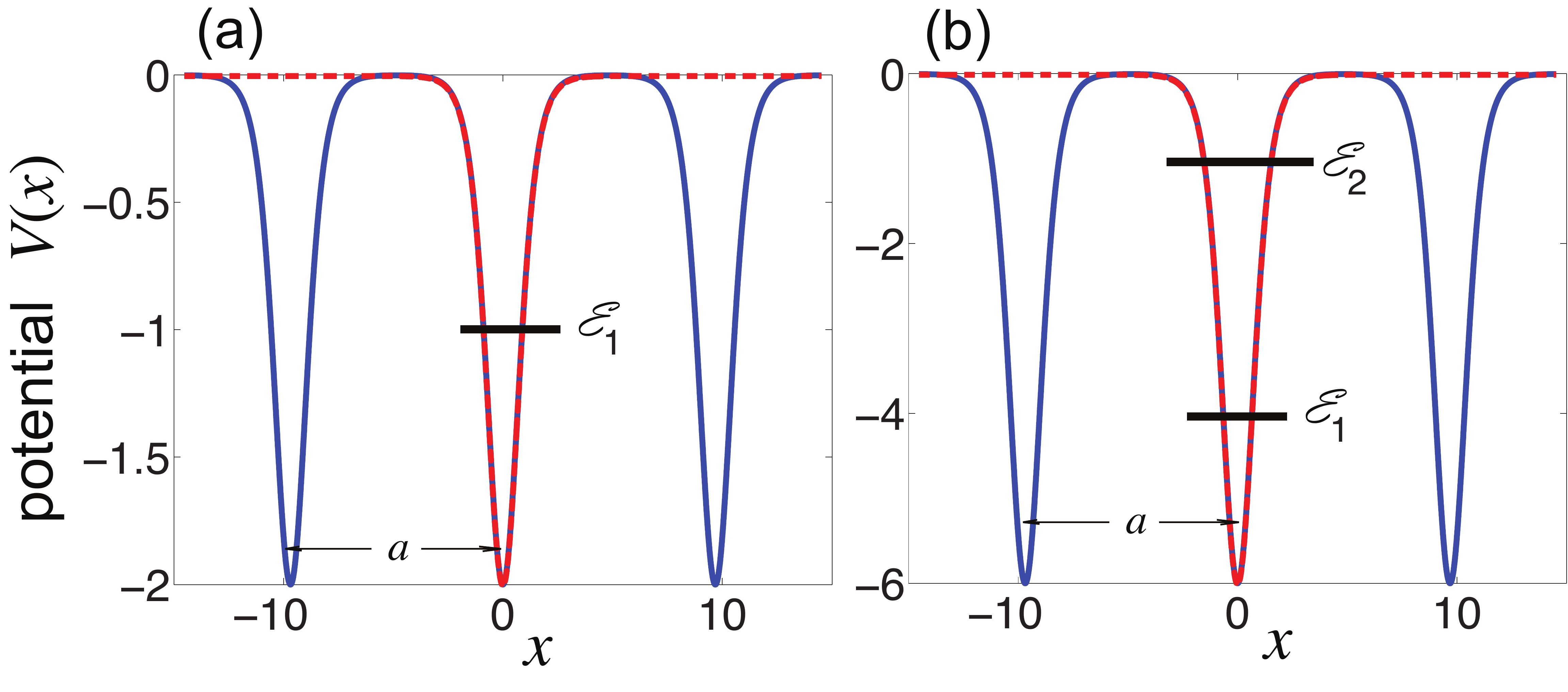}
    \caption{Behavior of the Lam\'e potential [Eq.(\ref{Lame})] for $m=0.999$ and for (a) $N=2$, and (b) $N=3$. The lattice period is $a=2 K(m) \simeq  9.682$. In each period, the potential reproduces the reflectionless P\"oschl-Teller potential well (dashed curves), which sustains $N$ bound states with energies $\mathcal{E}_1=-1$ in (a), and $\mathcal{E}_1=-4$, $\mathcal{E}_2=-1$ in (b).}
     \label{fig2}
\end{figure}
(see Fig.3). The P\"oschl-Teller potential well defined by Eq.(\ref{PTwell}) sustains $(N-1)$ bound states with energies $\mathcal{E}_1=-(N-1)^2$, $\mathcal{E}_2=-(N-2)^2$, ..., $\mathcal{E}_{N-1}=-1$. Therefore, in the Hermitian limit $\beta=0$ the periodic potential $V(x)$ gives rise to $(N-1)$ narrow tight-binding bands centered at around  $\mathcal{E}_1$, $\mathcal{E}_2$, ..., $\mathcal{E}_{N-1}$, with the additional continuum band $0<E< \infty$ which cannot be described within a tight-binding approximation [see panels (a) in Figs.4 and 5]. As $\beta$ is slightly increased above zero, the tight-binding bands describe near ellipsoidal curves in complex energy plane, according to the tight-binding analysis, while the continuum unbounded band describes an open curve which goes to infinity. This is shown in panels (b) of Figs.4 and 5 for the case $N=2$ and $N=3$, respectively. As $\beta$ is further increased, a cascade of band merging is observed [Fig.4(c) and Figs.5(c-e)], until above a critical value $\beta_c$ the energy spectrum is described by a single open curve unbounded at infinity, which converges toward the parabola of the free-particle problem in the large $\beta$ limit. i.e. to the curve with cartesian equation
\begin{equation}
{\rm Re}(E) \simeq V_0-\beta^2+ \left(  \frac{{\rm Im}(E)}{2 \beta} \right)^2
\end{equation}
[Figs.4(d) and 5(f)]. \par
\begin{figure}[t]
   \centering
    \includegraphics[width=0.50 \textwidth]{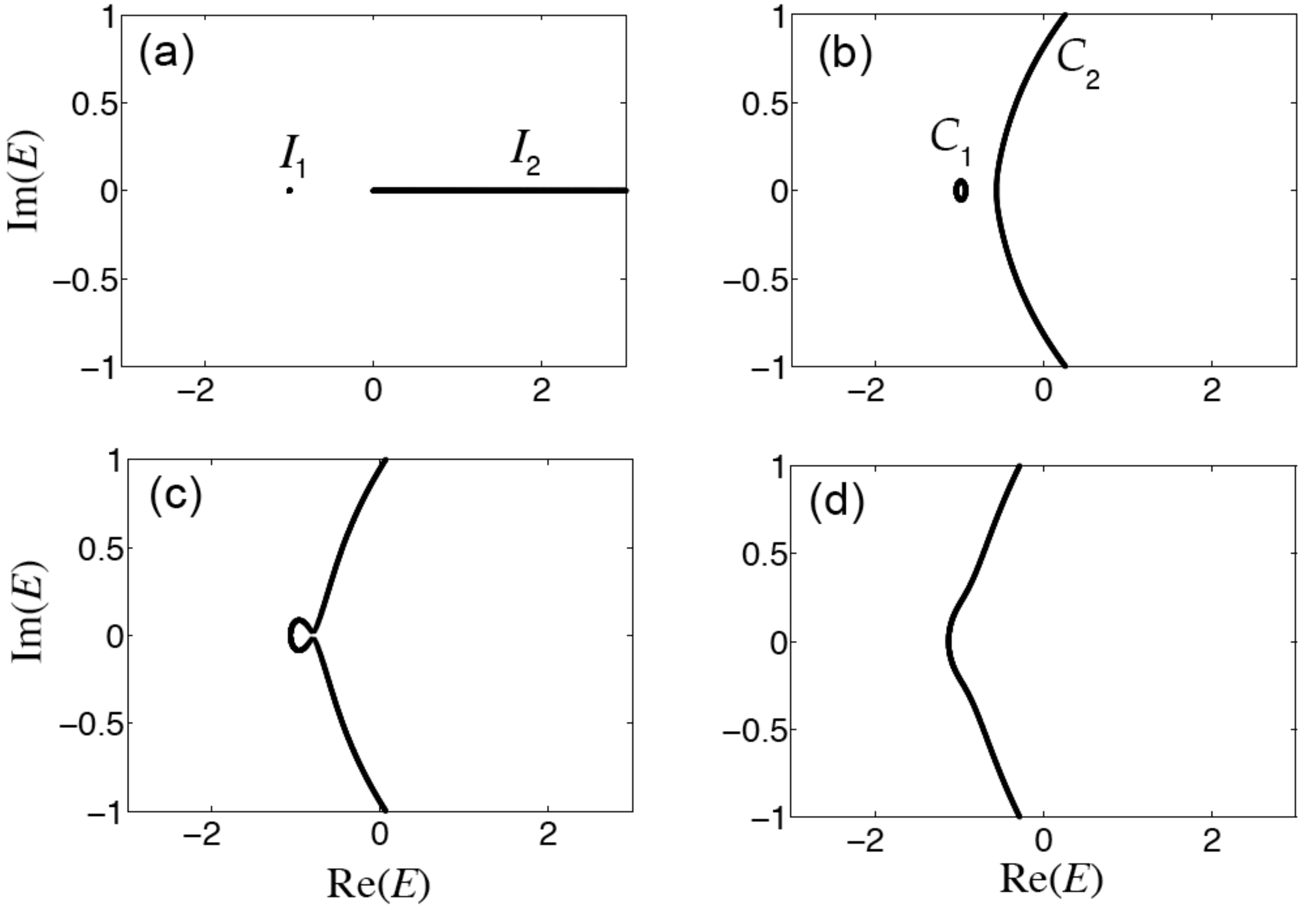}
    \caption{Energy spectrum under PBC of the Lam\'e potential (\ref{Lame}) for $m=0.999$, $N=2$ and for a few increasing values of the imaginary gauge field $\beta$. (a) $\beta=0$ (Hermitian limit), (b) $ \beta=0.55$, (c) $\beta=\beta_c \simeq 0.5963$, and (d) $\beta=0.7$. The critical value $\beta=\beta_c$, above which band merging occurs, is given by Eq.(\ref{critical}).}
     \label{fig4}
\end{figure}
\begin{figure}[t]
   \centering
    \includegraphics[width=0.50 \textwidth]{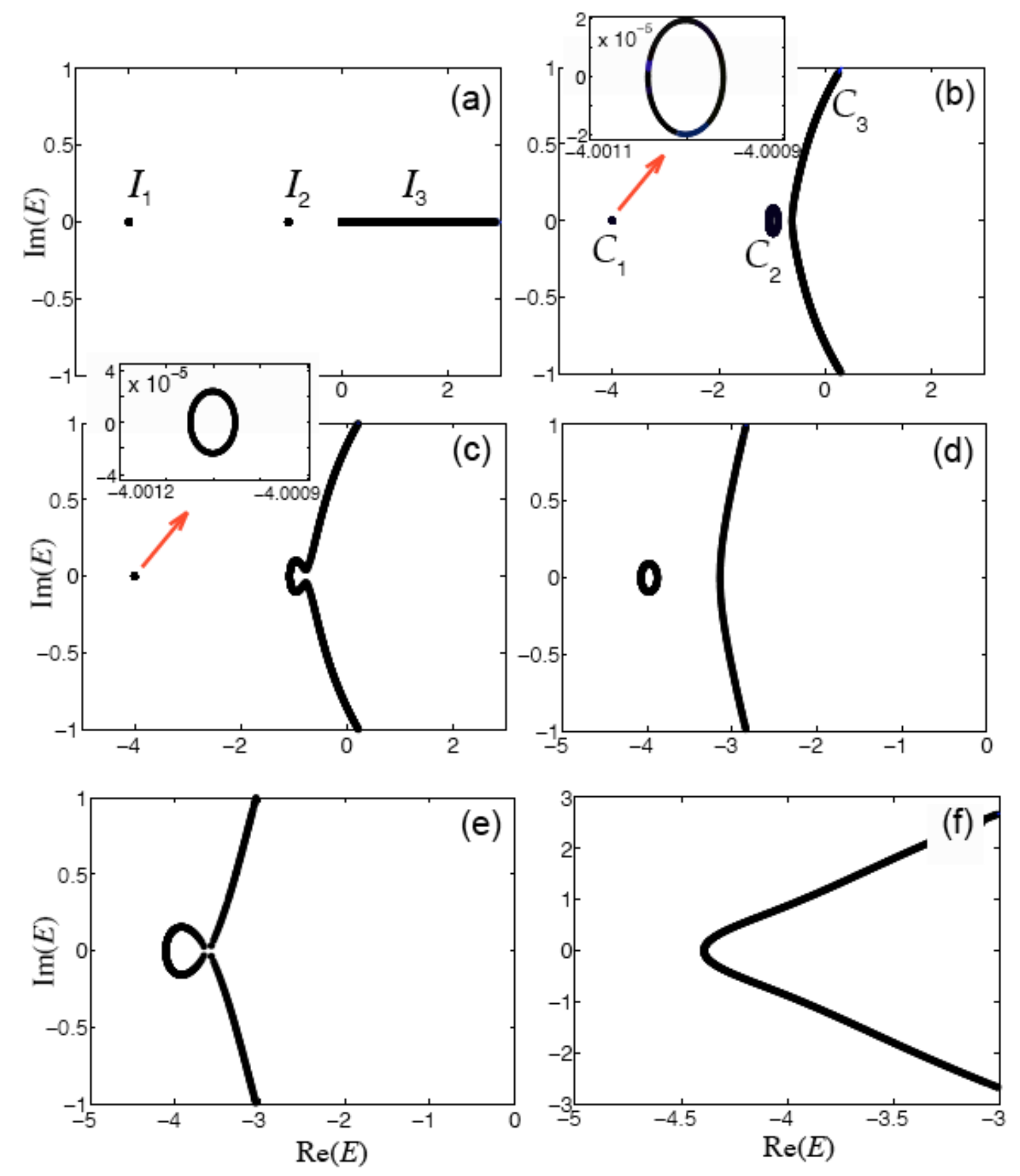}
    \caption{Energy spectrum under PBC of the Lam\'e potential (\ref{Lame}) for $m=0.999$, $N=3$ and for a few increasing values of the imaginary gauge field $\beta$. (a) $\beta=0$ (Hermitian limit), (b) $ \beta=0.48$, (c) $\beta=0.5$, (d) $\beta=1.35$, (e) $\beta=1.4$, and (f) $\beta=1.6$. The insets in (b) and (c) show an enlargement of the ellipsoid loop $\mathcal{C}_1$ of the lowest tight-binding band of the lattice, emanating from $I_1$.}
     \label{fig4}
\end{figure}
To clarify the band-merging effect and to calculate the critical value $\beta_c$, let us focus our attention to the $N=2$ case, which is amenable for a simple analytical study. For $N=2$, in the Hermitian limit $\beta=0$ the Lam\'e potential sustains two bands $I_1$ and $I_2$, with band edge energies given by Eq.(\ref{Eedges}) [see also Fig.4(a)]. For $m$ close to one, in the interval $-a/2 \leq x < a/2$ the potential $V(x)$ can be approximated by the reflectionless potential (\ref{PTwell}), so that using the supersymmetric properties of the P\"oschl-Teller potential $V_a(x)$ \cite{PT3} the Bloch wave functions $\psi(x)$ corresponding to the energy $E=k^2$ can be given in a simple form and read
\begin{equation}
\psi(x)=\left\{ -ik  + \tanh (x) \right\} \exp(i kx)
\end{equation}
with $k$ real. For a non-vanishing imaginary gauge field $\beta>0$, a formal solution to the eigenvalue equation $\hat{H}_{\beta} \psi(x)=E \psi(x)$ with energy $E=k^2$ is given by
\begin{equation}
\psi(x)=\exp(- \beta x) \left\{ -ik x + \tanh (x) \right\} \exp(i kx) \label{Ansat}
\end{equation}
where $k=k_R-ik_I$ can take rather generally complex values with real and imaginary parts $k_R$ and $-k_I$, respectively. When $\beta \neq 0$, to satisfy the PBC the allowed values of $k$ in complex plane should be taken such that
\begin{equation}
\left| \psi \left(x=-\frac{a}{2} \right) \right|=\left| \psi \left(x=\frac{a}{2} \right) \right| \label{BlochT}
\end{equation}
Since for $a \gg1$ one has $\tanh ( \pm a/2) \simeq \pm 1$, from Eqs.(\ref{Ansat}) and (\ref{BlochT}) one obtains
\begin{equation}
\exp [2 (\beta-k_I)a]=\frac{k_R^2+(1-k_I)^2}{k_R^2+(1+k_I)^2} \label{trasc}
\end{equation}
with corresponding energy
\begin{equation}
E=k^2=(k_R-ik_I)^2. \label{energy}
\end{equation}
For any given $-\infty < k_R < \infty$, Eq.(\ref{trasc}) is a transcendental equation for $k_I=k_I(k_R)$, with admits of up to three roots with  $k_I(-k_R)=k_I(k_R)$ and $k_I(k_R) \sim \beta$ as $|k_R| \rightarrow \infty$. For $\beta>\beta_c$, only one root of the transcendental equation (\ref{trasc}) is found. To calculate the critical value $\beta_c$, let us assume $k_R=0$, corresponding to a real and negative energy $E=-k_I^2$ according to  Eq.(\ref{energy}). Then Eq.(\ref{trasc}) can be solved for $\beta$, yielding
\begin{equation}
\beta=k_I+\frac{1}{a} \log \left|  \frac{1-k_I}{1+k_I}\right|. \label{beta}
\end{equation}
The behavior of $\beta$ versus $k_I$, as given by Eq.(\ref{beta}),  is shown in Fig.6. Clearly, for $\beta<\beta_c$, there are three allowed values of $k_I$, corresponding to the three real energies $E=-k_I^2$ where the two curves $\mathcal{C}_1$ and $\mathcal{C}_2$, emanating from bands $I_1$ and $I_2$, cross the real energy axis [Fig.4(b)]. On the other hand, for $\beta> \beta_c$ there is only one allowed value of $k_I$, corresponding to a single real energy belonging to the energy spectrum [the intersection of the open curve with the real energy axis in Figs.4(d)]. The critical value $\beta_c$, at which band merging occurs, is the relative maximum of the curve $\beta=\beta(k_I)$ of Fig.6 defined by Eq.(\ref{beta}), and can be readily calculated by letting $(d \beta / d k_I)=0$. This yields
\begin{equation}
\beta_c=\sqrt{1-\frac{2}{a}}+ \frac{1}{a} \log  \left( \frac{1-\sqrt{1-\frac{2}{a}}}{1+\sqrt{1-\frac{2}{a}}} \right). \label{critical}
\end{equation}
 \begin{figure}[t]
   \centering
    \includegraphics[width=0.45 \textwidth]{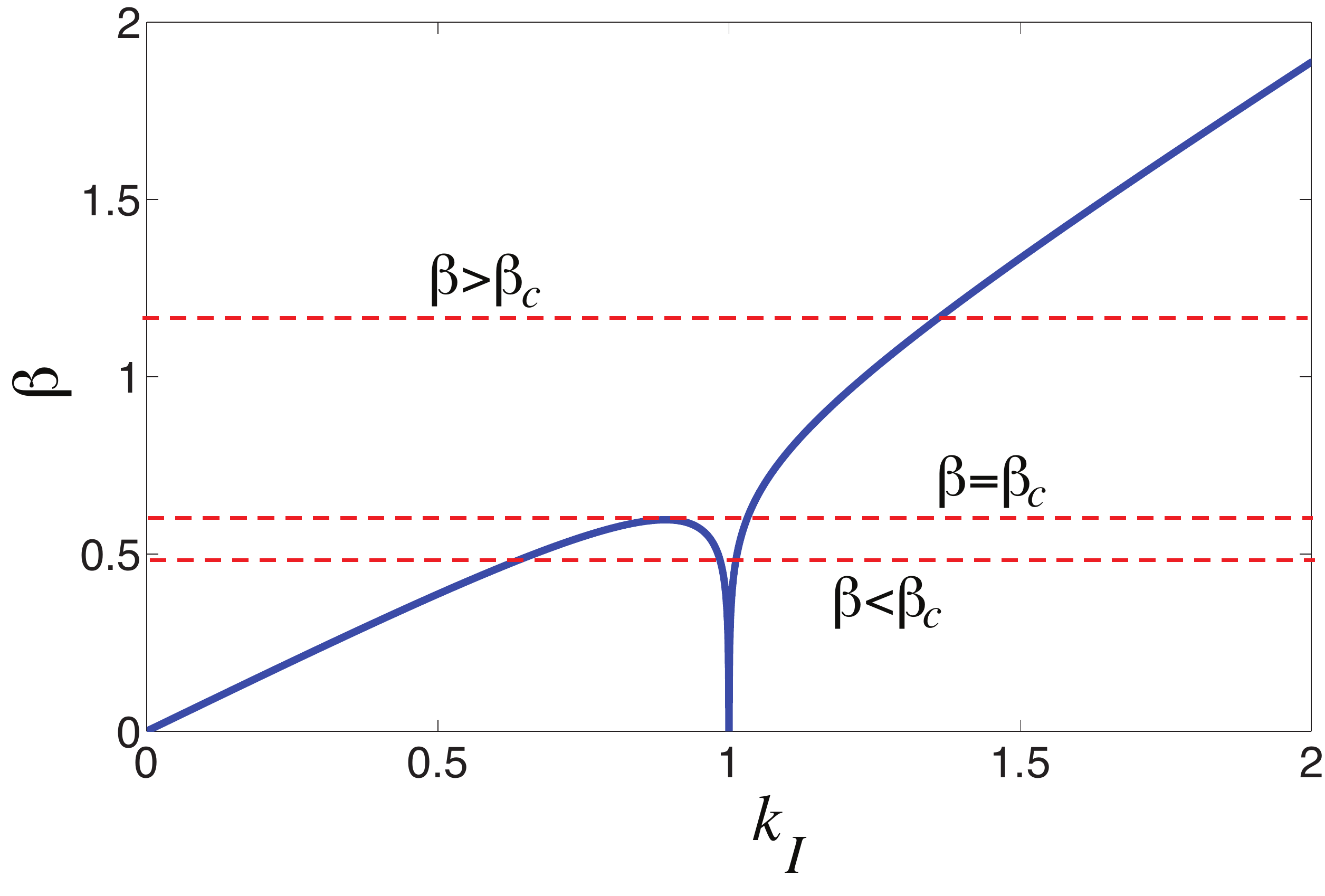}
    \caption{Behavior of the curve $\beta=\beta(k_I)$, defined by Eq.(\ref{beta}), for $m=0.999$, corresponding to a lattice period $a \simeq 9.682$. For $\beta<\beta_c$ there are three allowed values of $k_I$, while for $\beta> \beta_c$ there is only one allowed value of $k_I$.}
     \label{fig5}
\end{figure}

\subsection{The binary (double-well) potential}
As a second illustrative example, let us consider a potential $V(x)$ that describes a binary lattice and assume that in the interval (lattice period) $-a/2 \leq x < a/2$ the potential $V(x)$ is approximated by the reflectionless double-well potential \cite{DW1,DW2}
\begin{equation}
V(x) \simeq V_a(x)= 2( \sigma^2-1) \frac{\sigma^2+{\rm sech}^2 (x) \sinh^2 ( \sigma x)}{\left\{ \tanh(x) \sinh (\sigma x)- \sigma \cosh (\sigma x)  \right\}^2} \label{binary}
\end{equation}
with $\sigma>1$ and $a \gg 1$. The single potential well $V_a(x)$ sustains two bound states with energies $\mathcal{E}_1=-\sigma^2$ and $\mathcal{E}_2=-1$, spaced by $\Delta E= \sigma^2-1$ [Fig.7(a)]. The two energy levels  are almost degenerate in the $\sigma \rightarrow 1^+$ limit. In the periodic lattice,  two tight-binding mini bands are thus generated in this limit [bands $I_1$ and $I_2$ in Fig.7(b)], separated from a wide gap from the third continuum unbounded band [band $I_3$ in Fig.7(b)].

\begin{figure}[t]
   \centering
    \includegraphics[width=0.47 \textwidth]{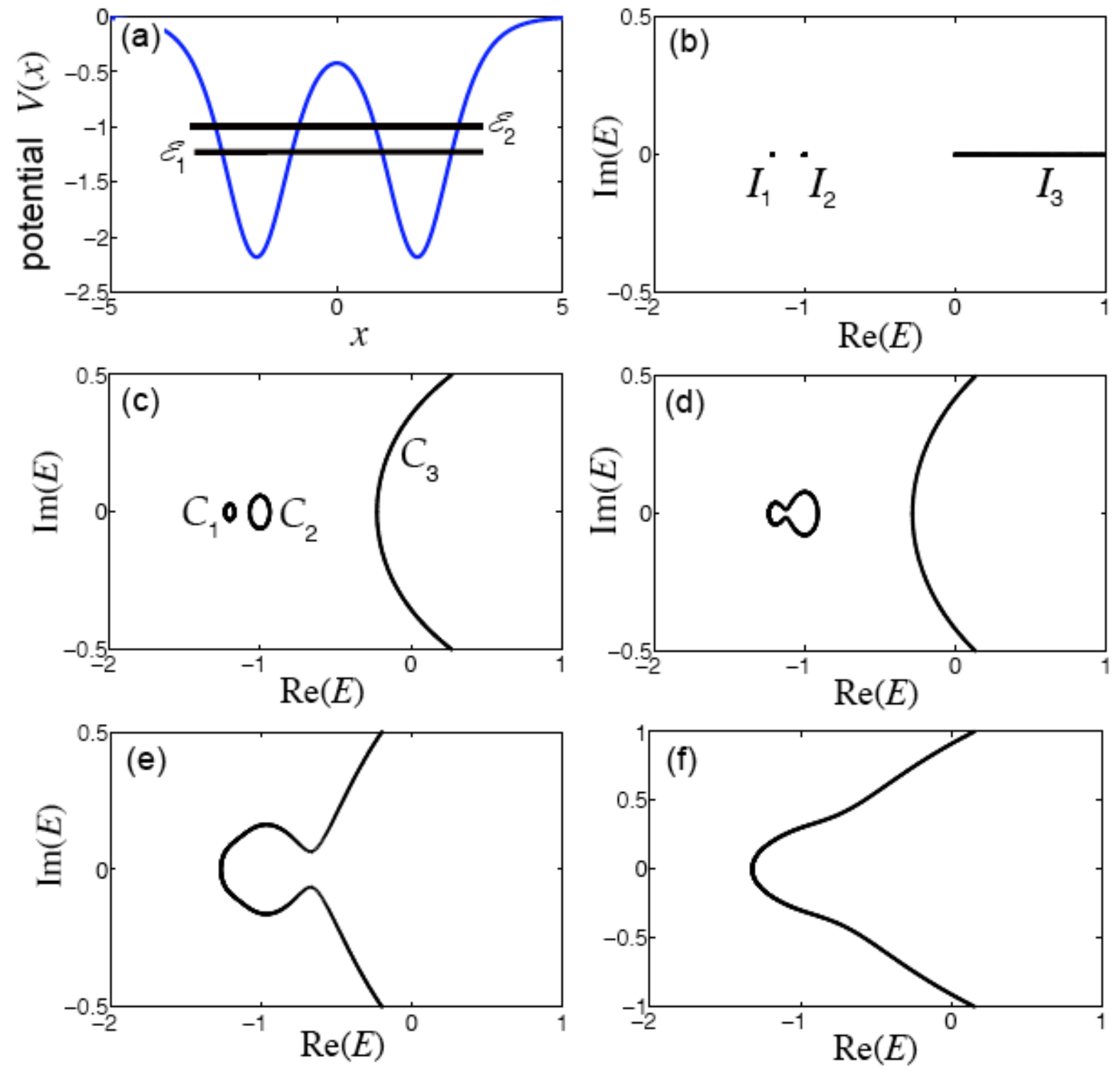}
    \caption{Energy spectrum under PBC of the binary potential (\ref{binary}) for $\sigma=1.1$ and lattice period $a=10$. (a) Behavior of the potential $V(x)$ in one lattice period. The double-well potential sustains two bound states with energies $\mathcal{E}_1=-\sigma^2$ and $\mathcal{E}_2=-1$ (solid horizontal lines). (b-f) Numerically-computed energy spectrum under PBC  for a few increasing values of the imaginary gauge field $\beta$: (b) $\beta=0$ (Hermitian limit), (c) $ \beta=0.28$, (d) $\beta=0.31$, (e) $\beta=0.4$,  and (f) $\beta=0.5$.}
     \label{fig4}
\end{figure}
As the imaginary gauge field $\beta$ is slightly increased above zero, the two mini bands are described by two closed loops (ellipsoids) [curves $\mathcal{C}_1$ and $\mathcal{C}_2$ in Fig.7(c)], until they merge in a single band with energy spectrum describing a closed loop and separated by the continuum band [Fig-.7(d)]. As $\beta$ is further increased, a second band merging is observed, with the spectrum being described in complex energy plane by a single open curve [Figs.7(e) and (f)]. We note that a tight-binding analysis could describe the band structure of the two mini bands, and their  merging as the imaginary gauge field is increased, however it cannot catch the entire band structure and the merging of the mini bands with the continuum band at large imaginary gauge fields. 

\subsection{The Mathieu (sinusoidal) potential}
The third illustrative example is provided by the Mathieu (sinusoidal) potential \cite{Slater,Mathieu}
\begin{equation}
V(x)=V_0 \cos (2 \pi x /a)
\end{equation}
which is not amenable for a tight-binding analysis, especially in the shallow potential limit $V_0 \ll (\pi /a)^2$.
As compared to the Lam\'e potential, in the Hermitian limit $\beta=0$ the sinusoidal potential displays an infinite number of bands separated by gaps, which become infinitesimally narrow as the energy increases.
Such gaps are related to the instability domains (resonance tongues) in the problem of parametric resonance \cite{tongue} and arise in correspondence of the resonance energies $E_n=(\pi/a)^2n^2$ ($n=1,2,3,...$).
\begin{figure}[t]
   \centering
    \includegraphics[width=0.50 \textwidth]{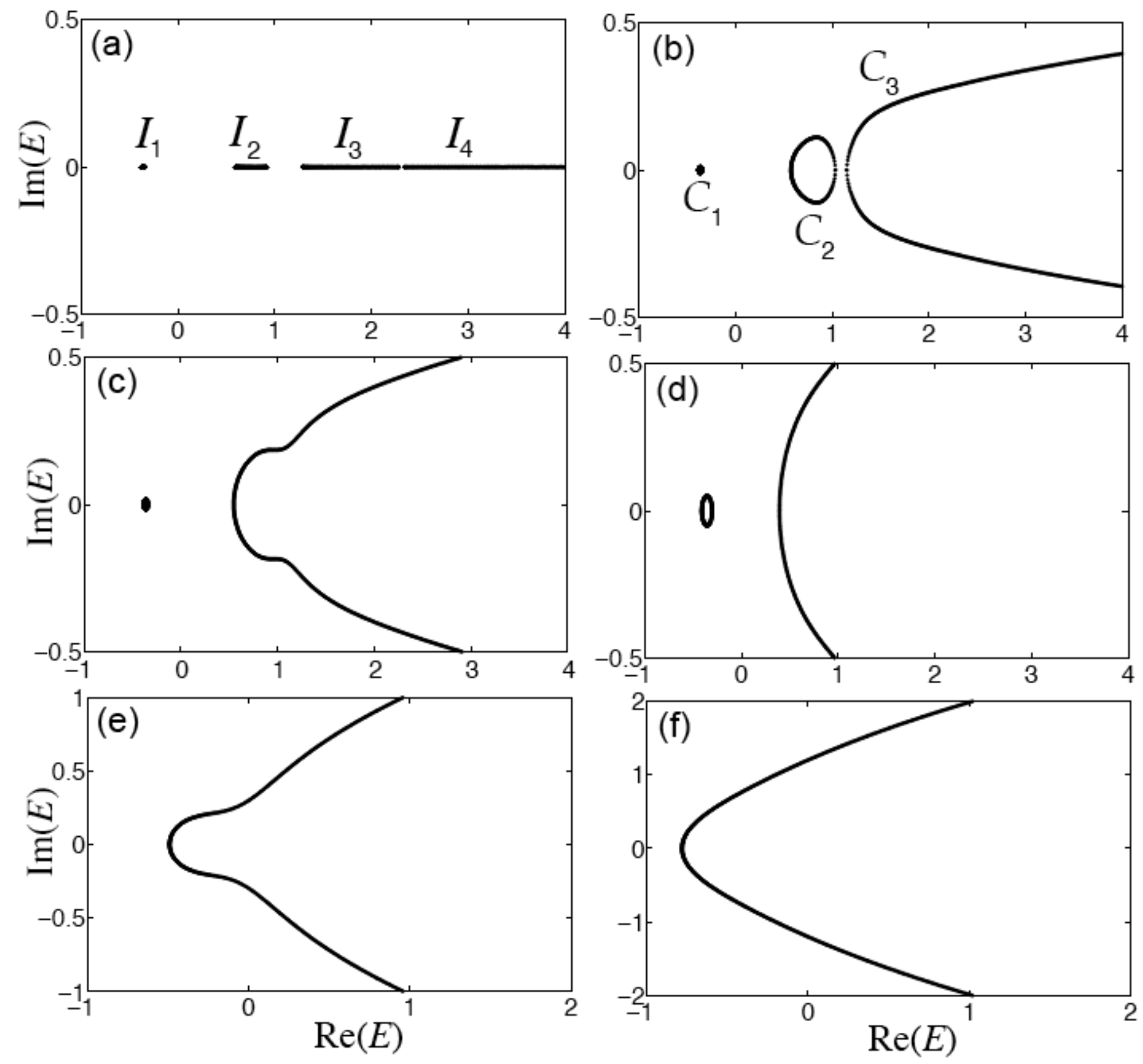}
    \caption{Energy spectrum under PBC of the Mathieu (sinusoidal) potential $V(x)=V_0 \cos (2 \pi x /a)$ for $V_0=1$, $a= 2 \pi$  and for a few increasing values of the imaginary gauge field $\beta$. (a) $\beta=0$ (Hermitian limit). Only three gaps, separating the bands $I_1$, $I_2$, $I_3$ and $I_4$, are clearly visible, while the other higher-energy gaps are too small to be visible. (b) $ \beta=0.1$. (c) $\beta=0.15$. (d) $\beta=0.3$. (e) $\beta=0.6$, and (f) $\beta=0.8$. A cascade of band merging is observed, until the energy spectrum in complex plane is described by a single open curve which is unbounded at infinity [panels (e) and (f)].}
     \label{fig4}
\end{figure}
 When a non-vanishing imaginary gauge field $\beta>0$ is applied, the bands separated by the many narrow gaps rapidly merge in a cascade process, until above a critical value $\beta=\beta_c$ all the bands are merged and the PBC energy spectrum in complex plane is described by an open curve that goes to infinity, approaching the parabolic curve (\ref{parabola}) in the large $\beta$ limit. The cascading process of band merging is illustrated in Fig.8. \par
The critical value $\beta_c$ becomes very small, with respect to $1/a$, in the shallow potential limit and  can be readily calculated by applying the nearly-free electron model \cite{Atland}, or equivalently the perturbation theory in the analogous problem of parametric resonance  \cite{tongue}. In fact, the critical value $\beta_c$ corresponds to the closing of the wider energy gap on the real energy axis, near the  energy $E_1=(\pi/a)^2$ of the first resonance tongue ($n=1$). {
In the nearly-free electron model and for a small value of the imaginary gauge field $\beta$, the Schr\"odinger equation (1) can be reduced to a Dirac-like equation with a non-Hermitian term arising from a non-vanishing value $\beta$.  In plane-wave (Bloch) basis, the Dirac Hamiltonian reads (details are given in Appendix B)
\begin{equation}
H(k)=
\left(
\begin{array}{cc}
E_1+2 k_0 (k-k_0-i \beta) & V_0/2 \\
V_0/2 & E_1-2 k_0 (k-k_0-i \beta)
\end{array}
\right)
\end{equation}
where we have set $k_0= \pi/a$ and $E_1=k_0^2$. For $\beta=0$, the eigenenergies $E_{\pm}(k)$ of $H(k)$ are given by the usual hyperbolic curves of the Dirac equation 
\begin{equation}
E_{\pm}(k)=E_1 \pm \sqrt{4k_0^2(k-k_0)^2+(V_0/2)^2}. \label{freee}
\end{equation}
which describe an avoided crossing of the bands near the edge $k=k_0$ of the Brillouin zone.} Such relations approximate the dispersion curves of upper and lower bands $I_2$ and $I_1$ near the wider energy gap at $E=E_1$ and for a Bloch wave number $k$ close to $k_0$.
Note that the width of the band gap is $V_0$, which corresponds to the difference $(E_+(k)-E_-(k))$ at the Bragg wave number $k=k_0$.
According to Eq.(\ref{complexification}), for a non-vanishing imaginary gauge field $\beta>0$ the energy dispersion curves are obtained from Eq.(\ref{freee}) by the substitution $k \rightarrow k-i \beta$. At $k=k_0$, the energies $E_{\pm}(k)$ are real and the width of the gap on the real axis reads 
\begin{equation}
(E_+(k_0)-E_-(k_0))=\sqrt{(V_0/2)^2-4 k_0^2 \beta^2}. \label{uffa}
\end{equation}
The critical value $\beta=\beta_c$ corresponds to the closing of the energy gap on the real axis. From Eq.(\ref{uffa}) one then obtains
\begin{equation}
\beta_c=\frac{V_0}{4 k_0}=\frac{V_0 a}{4 \pi}.
\end{equation}
{
Note that, at $k=k_0$ and $\beta=\beta_c$, the Dirac Hamiltonian (34) reduces to
\begin{equation}
H(k)=\left(
\begin{array}{cc}
E_1& 0  \\
0 & E_1
\end{array}
\right)+\frac{V_0}{2}
\left(
\begin{array}{cc}
-i & 1  \\
1 & i
\end{array}
\right)
\end{equation}
which is a defective $2 \times 2$ matrix, i.e. it does not have a complete basis of eigenvectors (the eigenvalue $E_1$ of $H(k)$ has an algebraic multiplicity of 2 but a geometric multiplicity 1). This means that, at the critical value $\beta=\beta_c$ the touching point of the two energy bands at $k=k_0$ is an exceptional point of the Hamiltonian. We note that a similar behavior, i.e. appearance of exceptional points and spectral singularities at the band merging points, is found in continuous models of complex crystals with parity-time ($\mathcal{PT}$) symmetry \cite{referee}.}

\section{Conclusions} In summary, we investigated the dependence of energy spectrum on boundary conditions and the NHSE in the framework of the one-dimensional continuous Schr\"odinger equation for a quantum particle in a periodic potential with an imaginary vector potential, beyond the usual tight-binding approximation. {The analysis reveals similarities and differences between the continuous and tight-binding models, that can be summarized as follows: (i) both models show the NHSE under OBC, i.e. the eigenstates are exponentially squeezed toward the lattice edges for any non-vanishing imaginary gauge field; (ii) in both models the NHSE is characterized  by a non-vanishing  point-gap topological winding number of the complex energy spectrum under PBC; (iii) unlike the tight-binding model, where the PBC energy spectrum is composed by closed curves in complex energy plane, in the continuous model the PBC energy spectrum always comprises an open curve, unbounded at infinity in a half complex plane and emanating from the high-energy states (nearly-free electronic states) of the Hermitian limit. At high values of the imaginary gauge field such an open curve describes the entire energy spectrum, as a result of a sequence of band merging illustrated in Fig.2; (iv) in both models the interior of the PBC energy spectrum corresponds to the energy spectrum of the non-Hermitian Hamiltonian under semi-infinite boundary conditions with localized edge states.}
 Such results have been illustrated by considering three significant examples of one-dimensional potentials, namely the Lam\'e potential supporting a finite number of gaps, the binary (double-well) potential supporting two tight.binding narrow bands, and the Mathieu (sinusoidal) potential, where the nearly-free electron model can be used to describe the band structure in the shallow potential limit.
Our results unravel the close connection between point-gap topology of the complex energy spectrum under PBC and the NHSE in continuous non-Hermitian systems beyond the usual tight-binding models, and is expected to stimulated further theoretical investigations on a rapidly developing area of research. For example, the analysis could be extended by considerning inhomogeneous (space-dependent) imaginary gauge fields, as well as other kinds of non-Hermitian terms in the continuous Schr\"odinger equation \cite{Pinotti}. Also, continuous models of non-Hermitian two-dimensional systems could be considered, where the second-order NHSE and corner states are observed within the tight-binding models \cite{r19,r52b,r54}.  Finally, the continuous non-Hermitian Schr\"odinger equation could be of relevance to investigate dual Hermitian systems in curved spaces \cite{curved}.

\acknowledgments
The author acknowledges the Spanish State Research Agency, through the Severo Ochoa
and Maria de Maeztu Program for Centers and Units of Excellence in R\&D (Grant No. MDM-2017-0711).

\appendix
\section{Energy spectrum of the semi-infinite lattice}
Let us consider the spectral problem in a semi-infinite lattice on the line $x \geq 0$, so that the Schr\"odinger equation (\ref{schroed}) should be supplemented with the semi-infinite boundary conditions (SIBC)
\begin{equation}
\psi(0)=0 \; , \; \; {\rm{max}} \lim_{x \rightarrow + \infty} |\psi(x)|< \infty. \label{semi}
\end{equation}
In this Appendix it is shown that, for a given value of the imaginary gauge field $\beta$, the energy spectrum of $\hat{H}_{\beta}$ under SIBC is provided by the domains in the interior of the PBC energy curves $\mathcal{C}_1$, $\mathcal{C}_2$,..., $\mathcal{C}_N$, depicted by the shaded areas in Fig.1. Moreover, for any energy $E_B$ strictly internal to such domains, the corresponding wave function is an edge state, exponentially localized at around $x=0$. To prove this statement, let $E_B$ be a complex energy strictly internal to one of such domains. Then there exists a value $\beta^{\prime} < \beta$ such that $E_B$ belongs to the PBC energy spectrum of $\hat{H}_{\beta^{\prime}}$: in fact, the shaded domains shown in Fig.1 can be obtained from the union of all the curves $\mathcal{C}_n$ of PBC energy spectra, emanating from the straight lines $I_n$, when the imaginary gauge field is adiabatically increased from 0 to $\beta$. Therefore, there exists a Bloch-type wave function $f(x)$ such that 
\begin{equation}
\hat{H}_{\beta^{\prime}}f(x)=E_B f(x)
\end{equation}
with
\begin{equation}
|f(x+a)|=|f(x)|.
\end{equation}
It can be readily shown that the wave function
\begin{equation}
\psi_1(x)= f(x) \exp [-(\beta-\beta^{\prime}) x]
\end{equation}
is formally an eigenfunction of $\hat{H}_{\beta}$ with the eigenenergy $E_B$, i.e.
\begin{equation}
\hat{H}_{\beta} \psi_1(x)=E_B \psi_1(x).
\end{equation}
Note that, since $\beta> \beta^{\prime}$, the wave function $\psi_1(x)$ exponentially decays toward zero as $x \rightarrow \infty$ like $\sim \exp[-(\beta-\beta^{\prime})x]$. The other linearly-independent solution to the second-order differential equation $\hat{H}_{\beta} \psi(x)=E_B \psi(x)$, namely
\begin{equation}
(\partial_x +\beta)^2 \psi(x)+-[E_B-V(x)] \psi(x)=0 \label{diffeq}
\end{equation} 
can be constructed from $\psi_1(x)$ as follows
\begin{equation}
\psi_2(x)= \psi_1(x) \int_0^x d \xi \frac{ \exp(-2 \beta \xi)}{\psi_1^2( \xi)}, 
\end{equation}
i.e.
\begin{equation}
\psi_2(x)= f(x) \exp [-(\beta-\beta^{\prime})x] \int_0^x d \xi \frac{ \exp(-2 \beta^{\prime} \xi)}{f^2(\xi)}.
\end{equation}
Clearly, also $\psi_2(x)$ exponentially decays to zero as $x \rightarrow + \infty$. The most general solution to the equation  (\ref{diffeq}) is thus given by
\begin{equation}
\psi(x)=A \psi_1(x)+ B \psi_2(x) \label{wf}
\end{equation}
with arbitrary constants $A$ and $B$. If we choose the constants such that $A \psi_1(0)+B \psi_2(0)=0$, i.e. $B=-A$, one has $\psi(x)=0$ for $x=0$ and $\lim_{x \rightarrow + \infty} \psi(x)=0$. This proves that $E=E_B$ belongs to the spectrum of $\hat{H}_{\beta}$ for the semi-infinite lattice and the corresponding wave function  (\ref{wf}) is an edge state. As $E_B$ approaches the domain boundaries $\mathcal{C}_1$,  $\mathcal{C}_2$, ..., $\mathcal{C}_N$, $\beta^{\prime} \rightarrow \beta^-$, the decay length of wave function diverges like $\sim 1/(\beta-\beta^{\prime})$ and $\psi(x)$ becomes an extended (Bloch-like) wave function, thus belonging again to the energy spectrum of $\hat{H}_{\beta}$ under the SIBC.

{
\section{The nearly-free electron model}
In the shallow potential limit, we can apply the standard nearly-free electron model to the Schr\"odinger equation (1) to describe the energy dispersion curves near a narrow energy gap. The analysis should be suitably extended to include a non-vanishing imaginary gauge field. For the sake of definiteness, we assume a shallow sinusoidal potential $V(x)=V_0 \cos (2 \pi x /a)$, discussed in Sec.III.C. Let $k_0= \pi /a$ be the Bloch wave number at the edge of the Brillouin zone and $E_1=k_0^2$ the energy of a free particle with momentum $k_0$ in the absence of the gauge field. For $V_0 \ll E_1$ and for a small value of $\beta$, we look for a solution to Eq.(1) of the form of two counter-propagating plane waves
\begin{equation}
\psi(x)= \psi_1(x) \exp(i k_0 x)+\psi_2(x) \exp(-i k_0 x)
\end{equation}
with carrier wave numbers $ \pm k_0$ and envelopes $\psi_1(x)$, $\psi_2(x)$ slowly varying on the spatial scale $\sim 1 / k_0$. Applying standard asymptotic methods and neglecting terms of order  $\sim \beta^2$, the envelopes $\psi_1(x)$ and $\psi_2(x)$ satisfy the coupled equations
\begin{eqnarray}
\epsilon \psi_1=-2 i k_0 \frac{\partial \psi_1}{\partial x}-2 i \beta k_0  \psi_1+ \frac{V_0}{2} \psi_2 \\
\epsilon \psi_2=2 i k_0 \frac{\partial \psi_2}{\partial x}+2 i \beta k_0 \psi_2+ \frac{V_0}{2} \psi_1.
\end{eqnarray}
where we have set $\epsilon=E-E_1$. For $\beta=0$, Eqs.(B2) and (B3) correspond to a one-dimensional Dirac equation for a massive and freely moving relativistic particle in the Weyl representation. The corresponding dispersion curves are given by Eq.(35) in the main text, which are readily obtained from Eqs.(B2) and (B3) after the Ansatz $\psi_{1,2}(x)= A_{1,2} \exp[i (k-k_0)x]$ with constant amplitudes $A_1$ and $A_2$. For a non-vanishing value of $\beta$, the Dirac equation becomes non-Hermitian and in Bloch basis is described by the $2 \times 2$ matrix $H(k)$ given by Eq.(34) in the main text, i.e.
\begin{equation}
E \left(
\begin{array}{c}
A_1 \\
A_2
\end{array}
\right)
= H(k) \left(
\begin{array}{c}
A_1 \\
A_2
\end{array}
\right).
\end{equation}

}

\end{document}